\documentclass{elsart}
\usepackage{ifpdf}
\usepackage{graphicx,amssymb,lineno}
\ifpdf
\usepackage[%
  pdftitle={Instructions for use of the document class
    elsart},%
  pdfauthor={Simon Pepping},%
  pdfsubject={The preprint document class elsart},%
  pdfkeywords={quark mass, chiral symmetry breaking, vacuum energy density},%
  pdfstartview=FitH,%
  bookmarks=true,%
  bookmarksopen=true,%
  breaklinks=true,%
  colorlinks=true,%
  linkcolor=blue,anchorcolor=blue,%
  citecolor=blue,filecolor=blue,%
  menucolor=blue,pagecolor=blue,%
  urlcolor=blue]{hyperref}
\else
\usepackage[%
  breaklinks=true,%
  colorlinks=true,%
  linkcolor=blue,anchorcolor=blue,%
  citecolor=blue,filecolor=blue,%
  menucolor=blue,pagecolor=blue,%
  urlcolor=blue]{hyperref}
\fi

\makeatletter
\def\elsartstyle{%
    \def\normalsize{\@setfontsize\normalsize\@xiipt{14.5}}
    \def\small{\@setfontsize\small\@xipt{13.6}}
    \let\footnotesize=\small
    \def\large{\@setfontsize\large\@xivpt{18}}
    \def\Large{\@setfontsize\Large\@xviipt{22}}
    \skip\@mpfootins = 18\p@ \@plus 2\p@
    \normalsize
}
\@ifundefined{square}{}{\let\Box\square}
\makeatother

\def\file#1{\texttt{#1}}

\usepackage{color}

\def\C{\textcolor{black}}

\newcommand{\be}{\begin{equation}}
\newcommand{\ee}{\end{equation}}
\newcommand{\bea}{\begin{eqnarray}}
\newcommand{\eea}{\end{eqnarray}}

\newcommand{\mbf}{\mathbf}

\begin{document}

\begin{frontmatter}
\title{Quark running mass and vacuum energy density in truncated Coulomb gauge QCD for five orders of magnitude of current masses}
\author{P. Bicudo}
\address{CFTP, Departamento de F\'{\i}sica, Instituto Superior T\'ecnico, \\
Avenida Rovisco Pais, 1049-001 Lisboa, Portugal}
\ead{bicudo@ist.utl.pt}

\begin{abstract}
We study in detail the effect of the finite current quark mass on chiral symmetry breaking, in the framework of truncated Coulomb gauge QCD
with a linear confining quark-antiquark potential. 
In the chiral limit of massless current quarks, the breaking of chiral symmetry is spontaneous. But for
a finite current quark mass, some dynamical symmetry breaking continues to add to the explicit breaking caused by the quark mass.  Moreover, 
using as order parameter the mass gap, i. e. the quark mass at
vanishing moment 
\C{
or the quark condensate, 
}
a finite quark mass transforms the chiral symmetry breaking from a phase transition into a crossover.  For the
study of the QCD phase diagram it thus is relevant to determine how the current quark mass affects chiral symmetry breaking. 
Since the current quark masses of the six standard flavours $u, \, d, \, s, \, c , \, b, \, t$ span over five orders of magnitude from 1.5 MeV to 171 GeV, we develop an accurate numerical method to study the running quark mass gap and the quark vacuum energy density from very small to very large current quark masses. 
\end{abstract}

\end{frontmatter}

\section{Introduction}

Chiral symmetry breaking has been studied in detail 
with chiral invariant {\em and confining} quark models
in the chiral limit of a 
vanishing current quark mass $m_0$.
For a finite current quark mass $m_0\neq 0 $,
studies exist with an approximate confinement
\cite{Munczek:1983dx,Jain:1991pk,Munczek:1991jb,Jain:1993qh}
or with a quadratic confining potential
\cite{Bicudo:1989sh,Bicudo:1989si,Bicudo:1989sj},
but very few studies have been performed 
\cite{linear4,LlanesEstrada:2004wr}
with an exactly linear confining potential.
Since the current quark masses of the six standard flavours $u, \, d, \, s, \, c , \, b, \, t$ span over five orders of magnitude from 1.5 MeV to 171 GeV, here we develop a new numerical method to study the quark mass gap and the quark vacuum energy density, with a linear exactly confining potential, from very small to very large quark masses.

Notice that even in the chiral limit of $m_0=0$, the quark has a constituent
running mass $m(p)$ function of the momentum $p$, solution of the mass 
gap equation (equivalent to the Schwinger-Dyson equation) for the quark.
Recently we have shown how to measure in the excited hadron spectra
the running mass $m(p)$
\cite{Bicudo:2009cr}.
The chiral invariant and confining quark models
have also been applied to phase studies at finite temperature 
$T$ and chemical potential $\mu$ 
\cite{Bicudo:1993yh,Battistel:2003gn,Antunes:2005hp,Glozman:2007tv,Guo:2009ma,Lo:2009ud,Kojo:2009ha,Nefediev:2009zzb}.
A finite quark mass is relevant both for the study of hadrons
which have been investigated for decades,
and for the study the QCD phase diagram which will be
explored in the future at RHIC, LHC and FAIR.
In the phase diagram, a finite current quark mass $m_0$
affects the position of the critical point between the 
crossover at low chemical potential $\mu$ and the phase transition
at higher $\mu$. Moreover the current quark mass
affects the QCD vacuum energy density $\cal E$, relevant
for the dark energy of cosmology.
This all occurs in the dynamical generation of the quark 
mass $m(p)$.
While the quark condensate $ \langle \bar \psi \psi \rangle$
is a frequently used order parameter for chiral symmetry breaking,
the mass gap, {\em i. e.    } the quark mass at vanishing
momentum $m(0)$ is another possible order parameter for
chiral symmetry breaking. However, due to technical difficulties,
$m(0)$ has not been computed in detail previously in confining
and chiral invariant quark models. 

Here we study in detail how a finite current quark
mass $m_0\neq 0 $ affects the dynamically generated quark
mass $m(p)$ .
We utilize the linear confining potential for the
quark-antiquark interaction, in the chiral quark model or Coulomb Gauge quark model,
including both confinement and chiral symmetry. 
While this model, in the framework Coulomb gauge Hamiltonian 
formalism is not yet full QCD, it is presently the only model able 
to include both the quark-antiquark confining potential and 
quark-antiquark vacuum condensation. 
Importantly, since our model is well defined and solvable, 
it can be used as a simpler model than QCD, and yet 
qualitatively correct, to address different aspects of hadronic physics. 
Is is adequate to study
the QCD phase diagram microscopically 
\cite{Bicudo:1993yh,Battistel:2003gn,Antunes:2005hp,Glozman:2007tv,Guo:2009ma,Lo:2009ud,Kojo:2009ha,Nefediev:2009zzb}.
We scan the current quark masses, from the light quarks to the
heavy quarks, computing the running quark mass $m(p)$
with detail, including the infrared limit of $m(0)$, {\em i. e.} the mass gap.

In Section II we derive in detail the mass gap equation. 
In Section III we review the numerical difficulties 
of this non-linear integral equation,
with cancelling infrared divergences.
We solve the mass gap equation with a 
new numerical method, dedicated to determine in detail the
difficult infrared region of small momentum $p \simeq 0$.
In Section IV we discuss our results and conclude.

\section{Our framework}

\subsection{Possible link to QCD.}

Our framework can be approximately derived from QCD,
in two different gauges.
In Coulomb gauge
$
\mathbf \nabla \cdot \mathbf A(\mathbf x, t)=0 
$
the interaction potential, 
has been derived by Lee,
\cite{TDLee}, 
and by Szczepaniak and Swanson
\cite{Szczepaniak:1995cw,Szczepaniak:1996gb}.
In the present study we address the quark fields only
and thus $V_I$ reduces to the quark part of the density-density term,
\def\kbar{\rlap\slash k}
\def\qbar{\rlap\slash q}
\def\pbar{\rlap\slash p}
\def\bl{{\bf l}}
\def\p{{\bf p}}
\def\a{{\bf a}}
\def\q{{\bf q}}
\def\k{{\bf k}}
\def\x{{\bf x}}
\def\y{{\bf y}}
\def\A{{\bf A}}
\def\B{{\bf B}}
\def\D{{\bf D}}
\def\bsig{{\bbox{\sigma}}}
\def\h{{1\over 2}}
\noindent
\begin{eqnarray}
V_I &=+& {1\over 2} g^2\int d\x d{\bf y} \, {\cal J}^{-1} \psi^{\dag}(\x) {\rm T}^a \psi(\x) 
\langle \x,a| \times
\nonumber \\
&&(\mathbf{\nabla} \cdot \D)^{-1} (-\mathbf{\nabla}^2)
(\mathbf{\nabla}\cdot \D)^{-1} | \y,b \rangle  {\cal J} \psi^{\dag}(\y) {\rm T}^b \psi(\y) \nonumber \\
\end{eqnarray}
The covariant derivative in the adjoint 
representation $\D = \mathbf{\nabla} - g \A$,
and ${\cal J} = \mbox{Det}[\mathbf{\nabla}\cdot \D]$ 
contribute to the density-density interaction, which
is expected to be confining in QCD.

Another approximate path from QCD to our model
considers the modified coordinate gauge of Balitsky 
\cite{Balitsky:1985iw},
$
\mathbf A(\mathbf 0, t)=0 \ , \, \mathbf x \cdot \mathbf A(\mathbf x, t)=0 
$
and in the interaction potential for the
quark sector,
\begin{eqnarray}
V_I=\int\, d^3x \left[ \psi^{\dag}( x) \;(m_0\beta -i{\vec{\alpha}
\cdot \vec{\nabla}} )\;\psi( x)\;+
{ 1\over 2} g^2 \int d^4y\, \
\right.
\nonumber \\
\overline{\psi}( x)
\gamma^\mu{\lambda^a \over 2}\psi ( x)  
\langle A_\mu^a(x) A_\nu^b(y) \rangle
\;\overline{\psi}( y)
\gamma^\nu{\lambda^b \over 2}
 \psi( y)  + \cdots \ 
\label{hamilt}
\end{eqnarray}
retains the first cumulant order, of two gluons
\cite{Dosch:1987sk,Dosch:1988ha,Bicudo:1998bz}
$
g^2 \langle A_\mu^a(x) A_\nu^b(y) \rangle
$
and this also results in a simple density-density harmonic effective 
confining interaction. As in QCD, this only has one scale.

Thus our framework is similar to an expansion of
the QCD interaction, truncated to the leading 
density-density term. 
Moreover, to address phenomenology
where the meson spectrum fits in linear Regge 
trajectories, 
one also needs to assume that the confining
quark-antiquark potential is a linear potential.
Notice that the short range Coulomb potential 
could also be included in the interaction, but 
here we ignore it since it only affects the quark 
mass through ultraviolet renormalization 
\cite{Bicudo:2008kc}, 
which is assumed to be already included in the 
current quark mass.
While this is not exactly equivalent to QCD,
our framework maintains three interesting
aspects of non-pertubative QCD,  
a chiral invariant quark-antiquark interaction,
\cite{Finger:1981gm,Orsay1,Orsay2,Orsay3,Orsay4,Kalinowski}
the cancellation of infrared divergences 
\cite{Bicudo:1989sh,Bicudo:1989si,Bicudo:1989sj}
and  a quark-antiquark linear potential
\cite{linear4,LlanesEstrada:2004wr,Szczepaniak:1995cw,linear1,linear2,linear3,Wagenbrunn:2007ie}.

\subsection{Deriving the mass gap equation}

We derive the mass gap equation,
where constituent quarks acquire 
the constituent mass $m(k)$
\cite{Bicudo_scapuz}
in the true and stable vacuum, 
solving the Schwinger-Dyson 
equation for the quark propagator,
\be
{\cal S}^{-1}(p) = 
{{\cal S}_0}^{-1}(p) 
\  - \
\begin{picture}(20,15)(0,0)
\put(0,0){\line(1,0){10}}
\put(22,0){\vector(-1,0){12}}
\put(10,-7){$_{k}$}
\put(7,18){$_{p-k}$}
\put(0,0){$\cdot$}
\put(1,4.4){$\cdot$}
\put(3,7){$\cdot$}
\put(5.6,9){$\cdot$}
\put(10,10){$\cdot$}
\put(14.4,9){$\cdot$}
\put(17,7){$\cdot$}
\put(19,4.4){$\cdot$}
\put(20,0){$\cdot$}
\end{picture}  
\ \ .
\ee
We utilize the truncated Schwinger-Dyson 
equation at the Rainbow level,
where the dotted line represents
the same density-density interaction $V_I$
resulting identically from the truncation of
Coulomb gauge QCD or of Balitsky gauge QCD. 
This leads to the same mass gap equation and quark 
dispersion relation obtained assuming a quark-antiquark $^3P_0$ 
condensed vacuum, computing the vacuum energy density
with the Hamiltonian, and minimizing the energy density.
The relativistic invariant Dirac-Feynman \cite{Orsay4}
propagators ${\cal S}$, 
can be decomposed in the quark and antiquark Bethe-Goldstone 
propagators
\cite{Bicudo_scapuz},
\begin{eqnarray}
\label{quarkpropagator}
&& {\cal S}(k_0,\vec{k})
= {i \over \not k -m +i \epsilon}
\\ \nonumber 
&=& {i \over k_0 -\sqrt{k^2 +m^2} +i \epsilon} \
\sum_su_su^{\dagger}_s \beta
- {i \over -k_0 -\sqrt{k^2 +m^2} +i \epsilon} \
\sum_sv_sv^{\dagger}_s \beta \ ,
\end{eqnarray}
where $m$ is the quark mass and where the quark spinor $u_s$ and antiquark spinor $v_s$ are,
\begin{eqnarray}
u_s({\bf k}) &=& \left[
\sqrt{ 1+S \over 2} + \sqrt{1-S\ \over 2} \widehat k \cdot \vec \sigma \gamma_5
\right]u_s(0)  
\nonumber \\
v_s({\bf k}) &=& \left[
\sqrt{ 1+S \over 2} - \sqrt{1-S \over 2} \widehat k \cdot \vec \sigma \gamma_5
\right]v_s(0)  
\ , 
\label{propagators}
\end{eqnarray}
where $ S=m /\sqrt{k^2+m^2} $ is a function of the quark mass.

Importantly, in the free propagator, the correct quark propagator in the non condensed vacuum,  
the quark mass $m$ is equal to the {\em current quark mass} $m_0 $. And it is this current
quark mass $m_0$ which effects in the current quark mass we study in great detail.
However when chiral symmetry breaking occurs,  $m$ is not determined from the onset.
In the physical vacuum, the {\em constituent quark mass} $m(k)$, 
is a function of the momentum, a dynamical solution of the mass gap equation.

Replacing the propagator
of eq. (\ref{propagators}) in the Schwinger-Dyson equation
and projecting it with the spinors, we get the mass gap equation
and the quark dispersion relation,
\bea
\label{2 eqs}
&&0 = u_s^\dagger(k) \left\{
k \widehat k \cdot  
\mbox{\boldmath \( \alpha \)}
 + m_0 \beta
-\int {d {k_0}' \over 2 \pi} {d^3 \mbf k' \over (2\pi)^3}
i \widetilde V(k-k') 
{\cal S}({k_0}',\vec{k'})
\right\} v_{s''}(k) \  \
\\ \nonumber 
&&E(k) = u_s^\dagger(k) \left\{k \widehat k \cdot 
\mbox{\boldmath \( \alpha \)} + m_0 \beta
-\int {d  {k_0}' \over 2 \pi} {d^3 \mbf k' \over (2\pi)^3}
i \widetilde V(k-k')  
{\cal S}({k_0}',\vec{k'})
\right\} u_s(k),
\label{SDE}
\eea
where the usual notation for Dirac matrices is assumed. 
Writing the running mass in terms of a sine and a cosine
of  $\varphi(k) = \arctan{ k \over m(k)} $, the {\em chiral angle},
\bea
S(k) &=& \sin \varphi(k) = {m( k) \over \sqrt{k^2 + m(k) ^2}} \ ,
\nonumber \\  
C(k) &=& \cos \varphi(k) = { k \over \sqrt{k^2 + m(k) ^2}} \ ,
\eea
the mass gap equation and the quark energy are,
\bea
0&=& + S(p) \,  B(p)- C(p)\, A(p) 
\label{mass gap}
\\
E(p)&=& + S(p) \,  A(p)+ C(p)\, B(p) 
\label{quark energy}
\eea
where the propagator functions $A$ and $B$,
respectively replacing the quark mass $m$ and
quark momentum $|\mathbf p|$ in the one-loop
dressed quark propagator of eq. (\ref{quarkpropagator}) are,
\bea
A(p)&=&  m_0 + {1 \over 2} \int {d^3 \mathbf k \over (2 \pi)^3}  \widetilde{V}( \mathbf p - \mathbf k)S(k) \ ,
\nonumber \\
B(p)&=&  p+ {1 \over 2} \int {d^3 \mathbf k \over (2 \pi)^3}  \widetilde{V}( \mathbf p - \mathbf k)(\hat p \cdot \hat k)C(k) \ .
\label{AB}
\eea
Equivalently to solve the non-linear integral mass gap equation
(\ref{mass gap}),
we can alternatively minimize the vacuum energy density per unit volume,
\be
{\cal E}= - {g \over 2} \int {d^3 \mathbf p \over (2 \pi)^3} 
 S(p) \left[ A(p) + m_0 \right] + C(p)  \left[ B(p) +p \right]  
\ee
where $g= N_f \, N_s \, N_c$ is the degeneracy factor counting the number of different but degenerate quarks.
$N_s=2$ is the number of spins and $N_c=3$ is the number of colours. 
$N_f$ is the number of degenerate flavours, but since each quark has a different
current quark mass $m_0$ one should compute separately the vacuum energy difference
for each quark flavour.

\subsection{The mass gap equation for a linear confining potential}

Notice that in the case of a linear potential, divergent in the
infrared limit of large $r$, the Fourier transform needs an infrared
regulator $\mu$.
A possible form of the linear potential,
\be
V(r)= - \sigma {e^{- \mu \, r} \over \mu} \simeq - {\sigma \over \mu} + \sigma \, r - { \sigma \mu \over 2 } \, r^2 +\cdots
\label{IRdivpot}
\ee
corresponds, in the limit of small infrared regulator $\mu$, to a model of linear confinement where the quark also has
an infinite binding energy $- {\sigma \over \mu}$. 
While other infrared regulations can be used for the linear potential, say $V(r)= \sigma r \, e^{- \mu \, r} $
which has no infrared divergent binding energy, the  infrared divergent constant of Eq. (\ref{IRdivpot}) is exactly 
cancelled in the mass gap equation by the  factor in the integrand $\bigl[S(k) C(p) 
 - S(p) C(k) \hat k \cdot \hat p \bigr] $.
The potential in Eq. (\ref{IRdivpot}) has a simple three-dimensional Fourier transform,
\bea
V(k) 
&=&  \sigma { - 8 \pi \over ( k^2 + \mu^2)^2   } \ , 
\eea
and this is the momentum space potential frequently utilized to account for linear confinement.

The integrals in the angular variable $\omega$ of Eq. (\ref{AB}) are,
\bea
&& \int_{-1}^1 d \omega 
{ -8 \pi \over ( k^2 + p^2 + 2 k p \omega + \mu^2)^2  }
= { - 16 \pi \over [(k-p)^2 + \mu^2] [(k+p)^2 + \mu^2]  } \ ,
\nonumber \\
&& \int_{-1}^1 d \omega
{ - 8 \pi \, \omega \over ( k^2 + p^2 + 2 k p \omega + \mu^2)^2  }
\\ \nonumber
&& \ \
= { - 16 \pi  \over (2 k p)^2 } 
\biggl\{ 
-{ 2 k p (k^2 + p^2 + \mu^2) \over [(k-p)^2 + \mu^2] [(k+p)^2 + \mu^2] } 
+ {1 \over 2} \log 
\left[ (k+p)^2 + \mu^2 \over (k-p)^2 + \mu^2\right] 
\biggr\} \  .
\eea
We find for the propagator functions $A$ and $B$,
\bea
A(p)&=&  m_0 - {\sigma \over p^2} \int_0^\infty {d k \over 2 \pi}  
I_A(k,p,\mu) S(k) \ ,
\nonumber \\
I_A(k,p,\mu) &=& { p k \over (p-k)^2 + \mu^2} 
- { p k \over (p+k)^2 + \mu^2} \ ,
\nonumber \\
B(p)&=&  p- {\sigma \over p^2}  \int_0^\infty {d k \over 2 \pi}  
I_B(k,p,\mu) C(k) \ ,
\nonumber \\
I_B(k,p,\mu) &=& 
 { p k \over (p-k)^2 + \mu^2} 
+ { p k \over (p+k)^2 + \mu^2}
+ {1 \over 2} \log  {(p-k)^2 + \mu^2 \over (p+k)^2 + \mu^2}\ ,
\eea
leading to the mass gap equation in the two equivalent forms of
a non-linear integral functional equation,
\bea
\label{massgabackbacktosincos}
0 &=& p S(p) - m_0 C(p) - { \sigma \over p^2}
\int_0^\infty {d k \over 2 \pi}  \, \bigl[
\\ \nonumber 
&&
I_A(k,p,\mu)\,  S(k) C(p) 
- I_B(k,p,\mu) \,  S(p) C(k) \bigr] \ ,
\eea
and of a minimum equation of the energy density, 
\bea
\label{energybacktosincos}
&& {\cal E} = { -g \over 2 \pi}\int_0^\infty  {dp \over 2 \pi}
 \biggl[
2p^3 C(p) + 2 p^2 m_0 S(p) + \sigma \times
\\ \nonumber 
&&
\int_0^\infty {d k \over 2 \pi} 
 I_A(k,p,\mu) \,   S(k) S(p) 
+ I_B(k,p,\mu) \,   C(p) C(k) \biggr]  \ .
\eea

Eq. (\ref{massgabackbacktosincos})  can be
rewritten as a fixed point equation for the quark mass function $m(k)$,
\bea
&& m(p) = m_0+ { \sigma \over p^3}
\int_0^\infty {d k \over 2 \pi}  {I_A(k,p,\mu)\, 
 m (k) p   -
I_B(k,p,\mu)\,
 m(p) k
\over \sqrt{k^2 + m(k)^2}}   \ .
\label{fixedpointeq}
\eea
Since the potential has an infinite constant independent of the mass,
in the variational equation we search for the extremum of the energy difference
\bea
\label{energydifference}
{\cal E}  - {\cal E}_0&=& { -g \over 2 \pi}\int_0^\infty  {dp \over 2 \pi}
 \biggl\{
2p^3 [ C(p) -C_0(p) ] 
+ 2 p^2 m_0 [S(p) -S_0(p)] 
\\ \nonumber 
&&
\ \ \ + \sigma \, \int_0^\infty {d k \over 2 \pi}   I_A(k,p,\mu) \,   \left[ S(k) S(p) - S_0(k) S_0(p) \right]
 \\  \nonumber
 && 
 \ \ \ \ \ \
+ I_B(k,p,\mu) \,   \left[ C(p) C(k) - C_0(k) C_0(p) \right]  \biggr\} \ .
\eea
where ${\cal E}_0$ is constant, and where we use the subindex $_0$ when the mass $m(p)$ is substituted
by the constant current mass $m_0$.
These two Eqs. (\ref{fixedpointeq}) and (\ref{energydifference})  constitute the
main object of our study.

\section{Solving the mass gap equation variationally 
with rational ansatze}

\subsection{Accurate numerical cancellation of infrared and ultraviolet divergences}

We use both Eq. (\ref{fixedpointeq}) and the minimization of  
 Eq. (\ref{energydifference}) to find the dynamical quark mass
 $m(k)$, but first we must regulate correctly their divergences.
The infrared divergences are present in the term
$ p k /[(p-k)^2 + \mu^2]$,
infrared divergent in the limit of a vanishing regulator $\mu \to 0$, 
which is present bot in the functions $I_A(k,p,\mu)$ and $I_B(k,p,\mu)$.
We must show that this infrared divergence is cancelled both in
the fixed point Eq. (\ref{fixedpointeq}) and in the energy density 
 Eq. (\ref{energydifference}).  
In what concerns the fixed point Eq. (\ref{fixedpointeq}), while the denominator 
diverges quadratically in $1/(k-p)^2$, the numerator $m (k) p   -  m(p) k$ is of 
order $(k-p)$ and thus the integrand diverges like $ 1 /(p-k)$ only, and it's integral
has finite principal value.
In the energy density difference the infrared divergence also cancels,
since the numerator common to the infrared  divergent terms,
\bea
&& S(p)S(k)-S_0(p)S_0(k)+C(p)C(k)-C_0(p)C_0(k)  
\nonumber \\
&=& - {1 \over 2} \times \bigl[{\varphi'}^2(p) - {\varphi'_0}^2(p) \bigr] (k-p)^2
\\ \nonumber 
&& - {1 \over 2} \bigl[ \varphi'(p) \varphi''(p) -  \varphi'_0(p)  \varphi''_0(p) \bigr] (k-p)^3 + o (p-k)^4
\eea
is then of order $(k-p)^2$.

In the ultraviolet part of the integrals, while each sub-term in the propagator
function integrands $I_A(k,p,\mu)$ and $I_B(k,p,\mu)$  is divergent, the actual 
sum is ultraviolet convergent, since in
the limit of large $k$ we have,
\bea
p \, I_A(k,p,0) &= & \sum_{n=1}^\infty 4 n { p^{2 n +1}  \over k ^{2 n} }
\ , 
\nonumber
\\
k \, I_B(k,p,\mu) & =  &  \sum_{n=1}^\infty 4 n \left({n+1 \over n + {\scriptstyle 1 \over \scriptstyle 2} }\right){ p^{2 n +1}  \over k ^{2 n} }
\ ,
\label{IAIBexpansions}
\eea
and thus the integrals in $k$ have ultraviolet integrable integrands decaying like 
$ p^3 /k^2$.
Also, the ultraviolet divergence in the kinetic terms of the energy density 
$2p^3  C(p) + 2 p^2 m_0 S(p)$ 
cancels due to the difference with $2p^3  C_0(p) + 2 p^2 m_0 S_0(p)$ 
if the mass difference $m(k) -m_0$ vanishes sufficiently fast in the
ultraviolet.

To address correctly both the infrared and ultraviolet divergences of
the integrals in $k$ of  Eqs.  (\ref{fixedpointeq}) and (\ref{energydifference}),
we divide the integral in two sections, the infrared one for
$ 0 < k < 2p $ and the ultraviolet one for $ 2p  < k < \infty$. In the infrared
region we compute the respective principal value, performing a symmetric sum 
centred in $k=p$, maintaining a very small regulator $ \mu$ just to cancel automatically
the contribution of $k=p$. In the ultraviolet region we use the change of variable
 \cite{linear1}
of Adler and Davis $ k \to x/ (1-x)$ with Jacobian $ 1 /(1-x)^2$
and with integration between $x = 2p /(1+2p)$ and 1. The change of variable
in the ultraviolet transforms, say an integral of rational functions $ 1/(1+k)^n$ into the
integral of polynomials $ (1-x)^{n-2}$ and thus it is adequate to the integral
of rational functions as we have here. Our numerical integrals in $k$
of a generic integrand $\cal I$ singular in $p$ are thus computed in the form,
\be
\int_0^\infty dk {\cal I}(k) = \mbox{P} \left[\int_0^{2 p} dk {\cal I}(k) \right]
+ \int_{2 p \over 1+ 2p} ^1 {dx \over (1- x )^2} {\cal I}\left( x \over 1 -x\right) \ ,
\label{numerical integration}
\ee
where each of the two numerical integrals can either be computed with
a rectangular, trapezoidal or Simpson sum or with the gauss quadrature method.

I one would discretize the quark mass $ m(p) $ in a series
of momenta $p_i$,  then the integrals of Eq. (\ref{numerical integration}) 
loses accuracy.  Finite differences would  require many interpolations, 
both in the infrared end and in the ultraviolet end, since the principal value requires requires that $k$ 
has many summation points smaller than $p$ and many other
larger than $p$. Moreover the correct integration of the integrand in
the ultraviolet large $k$ limit, where the integrand behaves like $ \left(p\over k\right)^2$, also
requires an integration extending beyond any value of $p$.
To solve this problem, we utilize a well defined ansatz for $m(k)$,
formally describe the parametrized as, 
\be
m(p)=m(p; c_1,c_2, \cdots, c_n) \ .
\label{formalanzats}
\ee
and this allows the numerical summation for the integrals in any point
of the integration domain.

In what concerns the numerical convergence to the solution for $m(k)$,
the fixed point equation is relatively unstable, particularly in the infrared region 
of $p \simeq 0$, when we are searching for the vacuum groundstate.
Notice that the mass gap equation had not only one, but an
infinite tower of solutions
\cite{Bicudo:2003cy}. The excited solutions dominate
the fixed point iteration, converging to the larger eigenvalue of the iterated matrix, 
because they minimize the denominator $\sqrt{k^2 + m^2}$.
Previously in the literature, the fixed point method was provided with extra stability
with two different methods, Adler and Davis 
used a cubic equation
and relaxation, to select the best solution
\cite{linear1}. 
Bicudo and Nefediev quasi-linearized
the fixed point equation and selected the desired eigenvalue,
corresponding either to the stable vacuum or to excited, false vacua
\cite{Bicudo:2003cy}. 
Thus they were able to find both the stable vacuum and the excited false vacua.
But these works have not yet determined in detail $m(0)$, since this demands
a very large numerical precision, and since most previous authors have focused
in computing the function $S(p)$ which in the infrared region is $S(0)=1$ 
regardless of the actual finite value of $m(0)$. 
Importantly, the present technique of minimizing the energy density directly 
tends to the right solution, which is the groundstate vacuum.

Interestingly, the variation of the ansatz parameters
$ c_1,c_2, \cdots, c_n$
of the energy density of the vacuum 
$
{\cal E}={\cal E}(c_1,c_2, \cdots, c_n)
$
utilized in minimization codes with gradient
method, utilizes the fixed point equation.
Computing the partial derivatives of the energy 
density we get,
\bea
{\partial {\cal E}(c_i)
\over
\partial c_i} 
&=& \int d p { \delta {\cal E}  \over \delta \varphi } 
{\partial \varphi 
\over \partial c_i }
\nonumber \\
&=& - { g \over 2 \pi} \int_0^\infty  {d p \over 2 \pi } 
(-2 p^2) {\cal R}(p;c_i)  \, {p \over p^2 + m^2(p;c_i) } { \partial m(p;c_i) \over \partial c_i} \ ,
\label{variationenergyparameters}
\eea
where ${\cal R}(p;c_i)$ is the right hand side of the mass gap
Eq.  (\ref{mass gap}),
\be
{\cal R}(p;c_i) = +S(p;c_i)B(p;c_i) - C(p;c_i)A(p;c_i) \ ,
\ee
utilized in the fixed point Eq. (\ref{fixedpointeq}).

\subsection{Choosing variational ansatze for the quark mass $m(p)$ }

To guide our choice of ansatze $m(p; c_1,c_2, \cdots, c_n)$, 
we first notice that the series expansion for $I_A$ and $I_B$ in Eq. (\ref{IAIBexpansions}), 
also apply when $k \leftrightarrow p$.
When replaced in the integral of 
the fixed point equation (\ref{fixedpointeq}) , the series
suggest that a series expansion of $m(p)$ should only 
have even terms, {\em i. e.} $m(p)$ should be a function of $p^2$.
$m(p)$ should also be a finite function since our integrals 
are finite.

In what concerns the asymptotic ultraviolet tail of the integral
of the fixed point equation (\ref{fixedpointeq}), there
are two different limits we can address. In the case
of a large current quark mass $m_0$, 
$ m_0 \over k^2 +{m_0}^2 $ interpolates between $1$ in the
infrared region of the integral and $  m_0 \over k $ in the
ultraviolet region of the integral. Using these
approximation, the components of the integral are analytical,
\C{
\bea
\label{analytical A B}
\int_0^\infty {d k \over 2 \pi} p \,  I_A(k,p,\mu) &=&  {p^3  \over 2 \, \mu} \ ,
\\
\int_0^\infty {d k \over 2 \pi}  k \, I_B(k,p,\mu) &=& {p^3  \over 2 \, \mu} \ ,
\nonumber 
\\
\int_0^\infty {d k \over 2 \pi} { p \over k}  I_A(k,p,\mu) &=& 
 {  p^2 \arctan { p \over \mu }\over \pi \, \mu} 
 \nonumber
 \\
  &=& { p^2 \over 2 \mu} - {p \over \pi} + o \left( \mu \right) \ ,
\nonumber 
\\
\int_0^\infty {d k \over 2 \pi}   I_B(k,p,\mu) &=& 
{- p \mu +   (p^2 + \mu^2)  \arctan { p \over \mu }\over \pi \, \mu} 
\nonumber \\
&=&
{ p^2 \over 2 \mu} - 2 {p \over \pi } +  o \left( \mu \right)
\ ,
\nonumber 
\eea
}
and thus adding the respective components we find that in the
infared dominated approximation the integrals cancel, while 
in the ultraviolet dominated approximation  the integral with   
$  m_0 \over k $ produces an ultraviolet behaviour of
\C{ 
$m(p) - m_0 \to { m_0 \sigma \over \pi \, p^2}$. 
}
Thus in the case of large $m_0$ we expect that $m(p) - m_0 $
decays in the ultraviolet like $ 1/ p^2$.

In the case where $m_0 \simeq 0$, assuming
then that in the large $p$ limit the dynamical quark 
mass vanishes sufficiently fast, the fixed point equation leads to,
\bea
m(p) 
&\to& 
 { \sigma \over p^3}
\int_0^\infty {d k \over 2 \pi}  {1 \over \sqrt{k^2 + m(k)^2}}  
 4 { k^2 \over p^2} m (k) p   
\nonumber \\
&\to&
 { 4 \sigma \over p4}
\int_0^\infty {d k \over 2 \pi}  {k ^2 m (k) \over \sqrt{k^2 + m(k)^2}}  
\eea
and, providing $m(p)$ decays faster than $1 / p^2$ for a
finite integral, this decays in the ultraviolet like $1 / p^4$.

The simplest possible ansatz for $m(p)-m_0$ we may have, function
of $p^2$, and encompassing both 
the behaviour in $1 / p^2$  for a large current quark mass 
$ m_0$ and the behaviour in $1 / p^4$ for a small current
quark mass is the rational function,
\bea
\label{ansatz4}
{\cal A}_3 (p ) &=& {1 \over c_0 + c_2 p^2 + c_4 p^4} \ .
\eea
\C{
This ansatz is a Pad\' e approximant, and to check whether our 
simple ansatz is sufficient, it is convenient to check that the next
Pad\' e approximant, 
}
a more flexible rational ansatz with
two more parameters,
\bea
\label{ansatz6}
{\cal A}_5 (p ) &=& {1 + n_2 p^2\over d_0 + d_2 p^2 + d_4 p^4+ d_6 p^6}
\eea
leads to the same result. 
In both ansatze of Eq. (\ref{ansatz4})
and Eq. (\ref{ansatz6}) we assume that the parameters are positive.
While Eq. (\ref{ansatz4}) is a decreasing
function, Eq. (\ref{ansatz6}) may have a different behaviour at the origin,
either with an initial increase, or with a steeper decrease, an thus 
it has room in it's parameter space to verify if the ansatz of 
Eq. (\ref{ansatz4}) is close to the correct solution of the mass gap
equation.

We also check numerically that ansatze with steeper infrared behaviours,
including in the denominator terms like a $c_1 k$ or a $c_{-1} /k$ would
not improve the solution since the best solution would have $c_1=c_{-1}=0$.
A better ansatz than ${\cal A}_3(p)$ is ${\cal A}_5(p)$, however the improvement
of the solution is very small, almost invisible to the naked eye in graphics.
The partial redundancy between the numerator and denominator parameters
of ${\cal A}_5(p)$ already slows the convergence to the minimum of the energy 
density $\cal E$. Thus an ansatz with more parameters than ${\cal A}_5(p)$
is not necessary.
The only ansatze we adopt here are the ones of the rational functions
${\cal A}_3(p)$  and ${\cal A}_5(p)$.

\subsection{One loop results for the large current mass $m_0 >> \sqrt \sigma $ limit }

We now compute the first iteration of the fixed point method starting with $m(k)=m_0$. 
In the simple case case of a constant mass $m(k)=m_0$, 
we can compute the integral in the fixed point
equation with a large precision. Defining the mass 
difference $ {\cal D}$,
\be
 {\cal D} (p) = m(p) - m_0 \ ,
\ee
we compute ${\cal D} (p) $ in the case a constant mass $m_0$ is used in the integrand,
\be
{\cal D} _0 (p) = { m_0 \sigma \over p^3}
\int_0^\infty {d k \over 2 \pi}  {I_A(k,p,\mu)\, 
 p   -
I_B(k,p,\mu)\,
 k   \over \sqrt{k^2 + m_0^2}} \, .
\ee
We get an accurate result for the integral $ {\cal D}_0 (p) $,
with a numerical integration decomposed according to Eq. (\ref{numerical integration}).

This provides a good quark mass solution to $m(p)=m_0+ {\cal D}(p)$
whenever the current quark mass is large, i. e. when $m_0 >> {\cal D}(p)$.
In that case the integral $ {\cal D}_0(p) $ only needs to be computed once, 
since this one-loop approximation is already excellent.

Moreover we can rescale in $m_0$, and then with a single computation
we get $m(p)$ for for any constant mass
$m_0$. Denoting $\tilde k = k / m_0$ and so forth we get, 
\bea
 {\cal D}_0 (p) &=& {  \sigma  \over  m_0 \tilde p^3}
\int_0^\infty {d \tilde  k \over 2 \pi}  {I_A (\tilde k, \tilde p, \tilde \mu)\, 
  \tilde p   -
I_B(\tilde k, \tilde p, \tilde \mu)\,
  \tilde k   \over \sqrt{ \tilde k^2 + 1}}   
\nonumber \\ 
&=& {\sigma \over m_0} {\cal F} ({p \over m_0})  \ ,
\nonumber \\
{\cal F}(p) &=& 
 {  1  \over  p^3}
\int_0^\infty {d k \over 2 \pi}  {I_A (k,  p, \mu)\, 
  p   -
I_B( k,  p,  \mu)\,
   k   \over \sqrt{  k^2 + 1}}   \ .
\label{D0}
\eea
Thus we only need to compute the dimensionless function ${\cal F}(p)$ and this
will produce the dynamical quark mass for any current quark
mass $m_0$ larger than the typical scale $\sqrt \sigma$ of our problem,
\be
m(p) \simeq m_0 + {\sigma \over m_0} {\cal F} ({p \over m_0})  \ .
\ee

\begin{figure}[t!]
\begin{center}
\includegraphics[width=0.8\columnwidth]{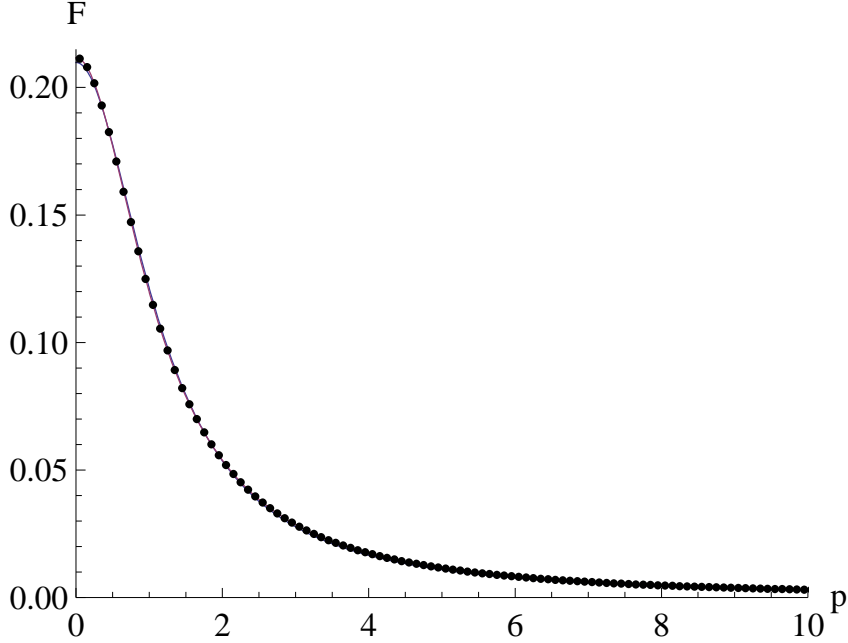}
\end{center}
\caption{Dimensionless  mass difference function ${\cal F}(\tilde p)$ computed in
the limit when the current quark mass $m_0$ is large. 
The dots show the numerical integral of Eq. (\ref{D0}) 
and the two almost overlapping solid lines show our fits with the 
two different rational ansatze of Eq. (\ref{ansatz4}) and Eq. (\ref{ansatz6}).
\label{massdiff1fop}}
\end{figure}

The solution of the integration, obtained with the simplest numerical
rectangular sum in both integrals, but needing $10^3$ integration points at least,
is represented in Fig. \ref{massdiff1fop}.

To be able to perform an accurate integration both in the infrared
and the ultraviolet, it is necessary to have an analytical function.
We obtain an analytical function by fitting the function ${\cal F} (p )$ ,
and we get an excellent fit already with the very simple ansatz
${\cal A}_3(p)$ of  Eq. (\ref{ansatz4}) for the parameter set,
\bea
c_0&=& 4.7664, \ c_2= 3.4762, \ c_4= 0.0000, \ 
\label{parset3}
\eea
with almost no graphically visible difference 
in Fig. \ref{massdiff1fop}
from the ansatz ${\cal A}_5(p)$
with two more parameters of Eq. (\ref{ansatz6}) 
for the parameter set,
\bea
d_0&=& 4.7250, \ d_2=4.9818, \ d_4= 0.8392, \
 d_6 = 0.0000, \ n_2= 0.2626,
\label{parset5}
\eea
and both fits confirm the $1/p^2$ decay of the mass in the ultraviolet,
for large current quark masses.

Now we can include the large quark limit scaling in $m_0$,
and also apply to the quark mass difference  ${\cal D}_0 (p )$
the ansatz $ {\cal A}_3 (p ) $ of Eq. (\ref{ansatz4}), and in this case the
ansatz parameters should be respectively,
\bea
{c_0}  = 4.7664 \, m_0 \ ,
{c_2}  = 3.4762 \,  m_0^{-1} \ ,
{c_4}  = 0.0000 \,  m_0^{-3} \ .
\label{parsofm0}
\eea


We further compute the vacuum energy density difference for a large  current quark  mass $m_0$. 
Although the mass difference ${\cal D}_0(p)$ in Eq. (\ref{D0}) decreases like
$ {m_0}^{-1}$ in the large  current quark  mass $m_0$ limit,
a dimensional analysis of ${\cal E} -{\cal E}_0$ shows
that possibly it does not vanish. Since the energy difference is finite
and independent of the infrared cutoff $\mu$, then our
only dimensionfull parameters are the string tension
$\sigma$ (scaling like a mass square) and the current quark mass $m_0$ (scaling like a mass).
Because the vacuum energy density per volume scales like a mass to the fourth power, 
a large mass expansion may have terms of the
form ${m_0}^4$, ${m_0}^2 \sigma$,  $\sigma^2$, $\sigma^3 {m_0}^{-2}, \cdots$
However the first term in this series clearly vanishes in the
vacuum energy difference ${\cal E} -{\cal E}_0$. Then the question
is what is the first non-vanishing term in this expansion.

\begin{table}[t!]
\begin{center}
\begin{tabular}{c| c c c c}
\hline 
$m_0$ & $ 10^4 {{\cal E} - {\cal E}_0 \over g} $ & $c_0$ & $c_2$ & $c_4$ \\
\hline
 $ 0.000100000 $ & $ -0.473813 $ & $ 6.24900 $ & $ 26.7910 $ & $ 17.5059 $  \\ 
 $ 0.000316228 $ & $ -0.478133 $ & $ 6.24287 $ & $ 26.7090 $ & $ 17.3392 $  \\ 
 $ 0.00100000  $ & $ -0.491581 $ & $ 6.22441 $ & $ 26.4168 $ & $ 16.8674 $  \\ 
 $ 0.00316228  $ & $ -0.532245 $ & $ 6.17597 $ & $ 25.4076 $ & $ 15.8086 $  \\ 
 $ 0.0100000    $ & $ -0.649468 $ & $ 6.01623 $ & $ 23.2517 $ & $ 12.0965 $  \\ 
 $ 0.0316228    $ & $ -0.958565 $ & $ 5.68682 $ & $ 18.5897 $ & $ 6.38068 $  \\ 
 $ 0.100000      $ & $ -1.62048 $ & $ 5.22374 $ & $ 12.5110 $ & $ 1.50433 $  \\ 
 $ 0.316228      $ & $ -2.58117 $ & $ 5.04452 $ & $ 7.22212 $ & $ 0.051756 $  \\ 
 $ 1.00000       $ & $ -3.27613 $ & $ 7.03428 $ & $ 3.06500 $ & $ 0.000000 $  \\ 
 $ 3.16228       $ & $ -3.46213 $ & $ 17.1365 $ & $ 1.04266 $ & $ 0.000000 $  \\ 
 $ 10.0000       $ & $ -3.47122 $ & $ 51.7243 $ & $ 0.336172 $ & $ 0.000000 $  \\ 
 $ 31.6228       $ & $ -3.39336 $ & $ 153.536 $ & $ 0.113085 $ & $ 0.000000 $  \\ 
\hline
\end{tabular}
\caption{The results for the ansatz parameters of ${\cal A}_3(p)$
\C{
for $m(p)-m_0$
}
obtained with our minimization code. All results are in dimensionless units of $\sigma=0.19$ GeV$^2=1$.
\label{table 1} }
\end{center}
\vspace{0.5cm}
\end{table}

\begin{table}[t!]
\begin{center}
\begin{tabular}{c| c c c c c c}
\hline 
$m_0$ & $ 10^4 {{\cal E} - {\cal E}_0 \over g} $ & $d_0$ & $d_2$ & $d_4$ & $d_6$ & $n_2$ \\
\hline
 0.000100000 &  -0.473908 &  6.170 &  31.9 &  22. &  14. &    0.5 \\ 
 0.000316228 &  -0.478221 &  6.163 &  32.3 &  23. &  15. &     0.6 \\ 
 0.00100000 &  -0.491647 &  6.147 &  32.2 &  24. &  16. &    0.6  \\ 
 0.00316228 &  -0.532266 &  6.09 &  33.1 &  30. &  19. &   0.9 \\ 
 0.0100000 &  -0.649397 &  5.9705 &  29.582 &  28.37 &  12.83 &    0.871 \\ 
 0.0316228 &  -0.958497 &  5.7202 &  25.358 &  32.40 &  7.4897 &    1.301 \\ 
 0.100000 &  -1.624761 &  5.369 &  11.875 &  5.05 &  0.0000 &    0.149 \\ 
 0.316228 &  -2.59104 &  5.467 &  6.720 &  2.08 &   0.0000 &  0.205 \\  
 1.00000 &  -3.27722 &  7.5349 &  4.4398 &  0.8105 &  0.0000 &    0.2546 \\ 
 3.16228 &  -3.46240 &  16.665 &  1.2612 &  0.0105 &  0.0000 &    0.0104 \\ 
 10.0000 &  -3.47148 &  49. &  0.6 &  0.00 &  0.0000 &    0.00 \\  
 31.6228 &  -3.76155 &  92. &  0.2 &  0.0000 &  0.0000 &  0.0000 \\
\hline
\end{tabular}
\caption{\label{table 2} The results for the ansatz parameters of ${\cal A}_5(p)$ 
\C{
for $m(p)-m_0$
}
obtained with our minimization code.  All results are in dimensionless units of $\sigma=0.19$ GeV$^2=1$.}
\end{center}
\vspace{0.5cm}
\end{table}

To answer this question, we expand the vacuum energy density in a $\sigma \over {m_0}^2$ series,
similar to the variation in Eq. (\ref{variationenergyparameters}), but now
based in the expansion of the quark dynamical mass,
\be
m(p) = m_0 + {\sigma \over {m_0}} m_1 \left(p \over m_0 \right) 
+ {\sigma^2 \over {m_0}^3} m_2 \left(p \over m_0 \right) + \cdots
\ee
and then we get 
\bea
 {\cal E}&=& 
{ -g \over 2 \pi}\int_0^\infty  {dp \over 2 \pi}
 \biggl\{ 2 p^2
 \biggl[ p C_0(p) + m_0 S_0(p) 
+ {  p \delta C_0(p)   + m_0 \delta S_0(p) \over \delta m_0  }  \left( m(p) -m_0 \right)  
\nonumber \\ &&
\hspace{2.5cm}
+{ 1 \over 2} { p \delta^2 C(p) + m_0 \delta^2 S_0 (p) \over \delta {m_0} ^2   } \left( m(p) -m_0 \right)^2 
+ \cdots \biggr]
\\ &&
+ \sigma \, \int_0^\infty {d k \over 2 \pi}  I_A(k,p,\mu) \,   \biggl[ S_0(p) S_0(k)  
 + 2 { \delta S_0(p) \over \delta m_0  }S_0(k )  \left( m(p) -m_0 \right)  + \cdots   \biggr]
\nonumber \\ \nonumber  && 
+ I_B(k,p,\mu) \,  \biggl[ C_0(p) C_0(k)  
+ 2 { \delta C_0(p) \over \delta m_0  }C_0(k )  \left( m(p) -m_0 \right)  + \cdots   \biggr]
  \biggr\} \ .
\eea
The zeroth order term vanishes when we perform the difference of the
vacuum energy densities ${\cal E}-{\cal E}_0$. Then the first
order term in the kinetic energy density also vanishes since the kinetic
energy density is minimized by $m(p)=m_0$.
Thus the leading term is of second order in the kinetic energy density,
and of first order in the potential energy density, 
and both are proportional to $ \sigma^2$. 
Using the intermediate steps,
\bea
&& { p \delta^2 C_0(p) + m_0 \delta^2 S_0 (p) \over \delta {m_0} ^2   } =
- {p^2 \over \left( p^2 + {m_0}^2 \right)^{3/2}} \ ,
\\ 
\nonumber 
&&  \sigma \int_0^\infty {d k \over 2 \pi}  \biggl[
{2  \delta S_0(p) \over \delta m_0  }  I_A(k,p,\mu) \,      S_0(k)  + I_B(k,p,\mu) \,   C_0(k)   \biggr]
{2  \delta C_0(p) \over \delta m_0  } 
\\ 
\nonumber 
&&  
\hspace{4.1cm}
 = 2 {p^4 \over \left( p^2 + {m_0}^2 \right)^{3/2}} {\sigma \over m_0} D_0 \left( p \over m_0 \right) \ ,
\eea
and changing variable to the dimensionless $ \tilde p $, we get,
\bea
{\cal E}-{\cal E}_0 
&=& \sigma^2   { -g \over 2 \pi} \int_0^\infty  {d \tilde p \over 2 \pi} 
{ \tilde p ^4  [  {\cal F} (\tilde p  )]^2 \over ( \tilde p^2 + 1 )^{7 \over 2} }
  + o\left(  \sigma^3 \over {m_0}^2 \right) . 
\eea
Finally using the ansatz $ {\cal A}_3 (p ) $ of Eq. (\ref{ansatz4}),
with the parameter set of Eq. (\ref{parset3})
this results in
\bea
{\cal E}-{\cal E}_0 
&\simeq& - - 3.47701 \times 10^{-4}  \, \sigma^2 \, g \ .
\label{enerofm0}
\eea

\begin{figure}[t!]
\includegraphics[width=.5\columnwidth]{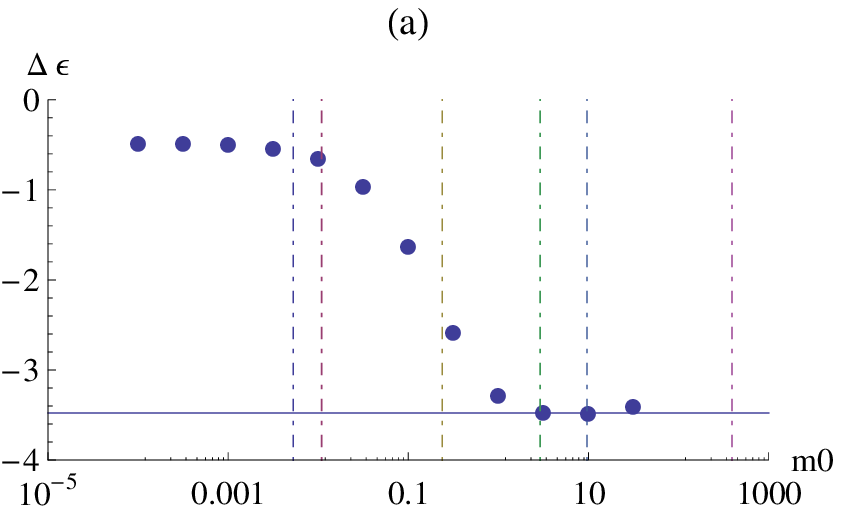}
\hspace{0.1cm}
\includegraphics[width=.5\columnwidth]{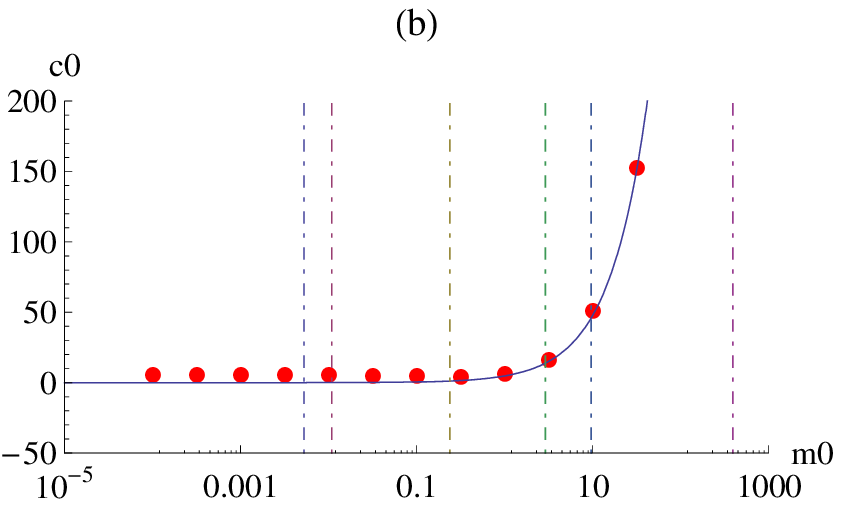}
\\
\\
\vspace{0.0cm}
\includegraphics[width=.5\columnwidth]{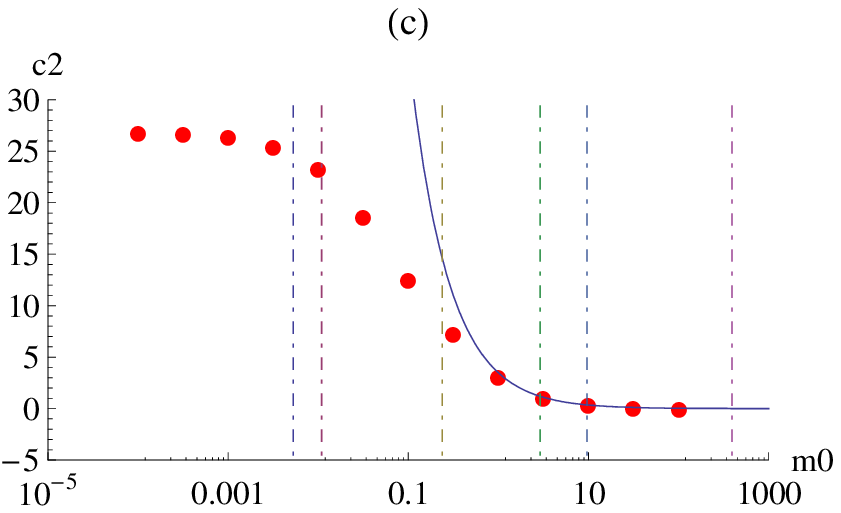}
\hspace{0.1cm}
\includegraphics[width=.5\columnwidth]{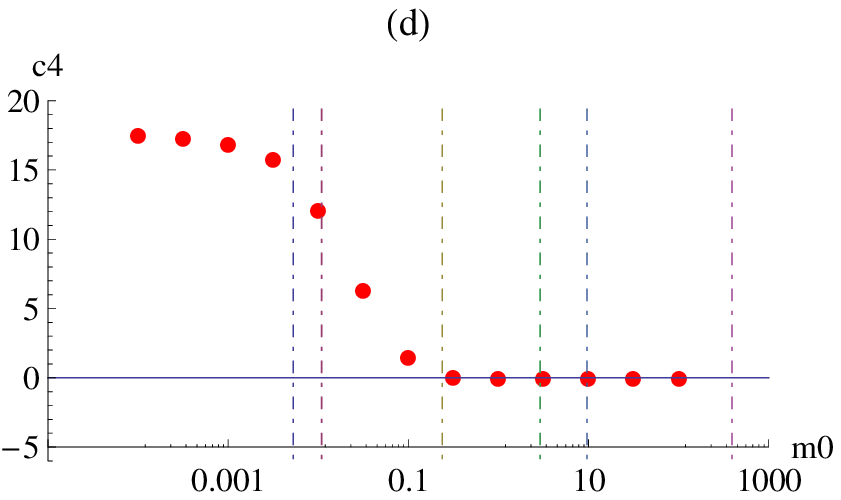}
\\
\caption{
Plots of our numerical solution with the ansatz ${\cal A}_3(p)$,
as a function of the current quark mass $m_0$: 
(a) the vacuum energy density shift ${\cal E} -{\cal E}_0$,
(b) parameter $c_0$ , 
(c) parameter $c_2$ ,   
(d) Parameter $c_4$.
The dots show our numerical solution, the solid line is the large $m_0$
limit obtained with Eqs. (\ref{parsofm0}) and  (\ref{enerofm0}), and
the vertical dot-dashed lines represent the
current masses of the quarks $u$, $d$, $s$, $c$, $b$, $t$.
\label{fig_solutionmodel4}}
\end{figure}

\begin{figure}[t!]
\includegraphics[width=.5\columnwidth]{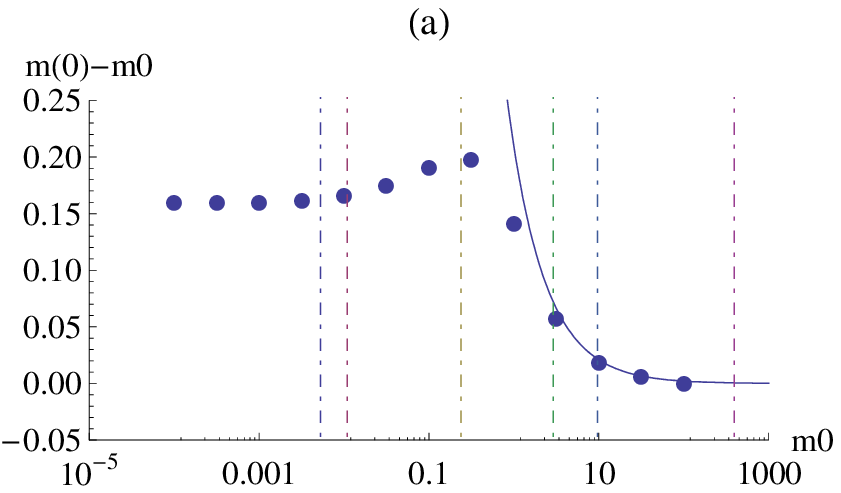}
\hspace{0.1cm}
\includegraphics[width=.5\columnwidth]{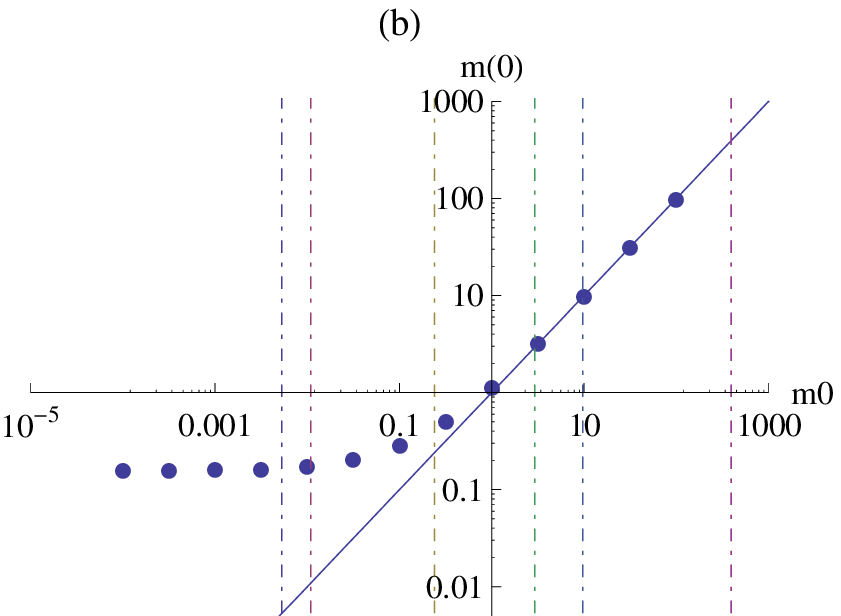}
\\
\caption{
Plots of the mass gap as a function of the current quark mass $m_0$: 
(a) the mass gap difference $m(0)-m_0$, measuring the amount
of generated dynamical mass, 
(b) the mass gap $m(0)$ . 
The dots show
our numerical solution, the solid line is the leading order 
obtained when $m_0 \to \infty$, and the vertical dot-dashed lines represent the
current masses of the quarks $u$, $d$, $s$, $c$, $b$, $t$.
Notice that the dynamical mass generation has a maximum
for finite quark masses close to the strange quark mass. 
All results are in dimensionless units of $\sigma=0.19$ GeV$^2=1$. 
\label{fig_massgap}}
\end{figure}

\section{Results}

Utilizing our ansatze of Eq. (\ref{ansatz4})  and Eq. (\ref{ansatz6}), 
we may now compute with great accuracy the
integrals of Eq. (\ref{energydifference}) which now are a function of the ansatz parameters,
and apply a standard minimizing code to determine the optimal
parameters.
We use 1000 $\times$ 1000 integration points and up
to 40 decimal digits in order to be able to find a convergence of the method
in the case of the ansatz ${\cal A}_5(p)$, due to the partial redundancy of the parameters.  
Then we also minimize the energy starting from different randomly generated initial values for the parameters. 
In Tables \ref{table 1} and \ref{table 2} we only show the digits which are stable, 
in the sense that they do not depend on the initial values.
Notice that the energy density obtained with the two different ansatze differ only
by a few per mil and that the ansatz of Eq. (\ref{ansatz6}) already exhibits some
redundance in the parameters. This shows that the ansatz of Eq. (\ref{ansatz4})  is 
already quite accurate for the parametrization of the quark mass $m(p)$.

\begin{figure}[t!]
\begin{center}
\includegraphics[width=0.8\columnwidth]{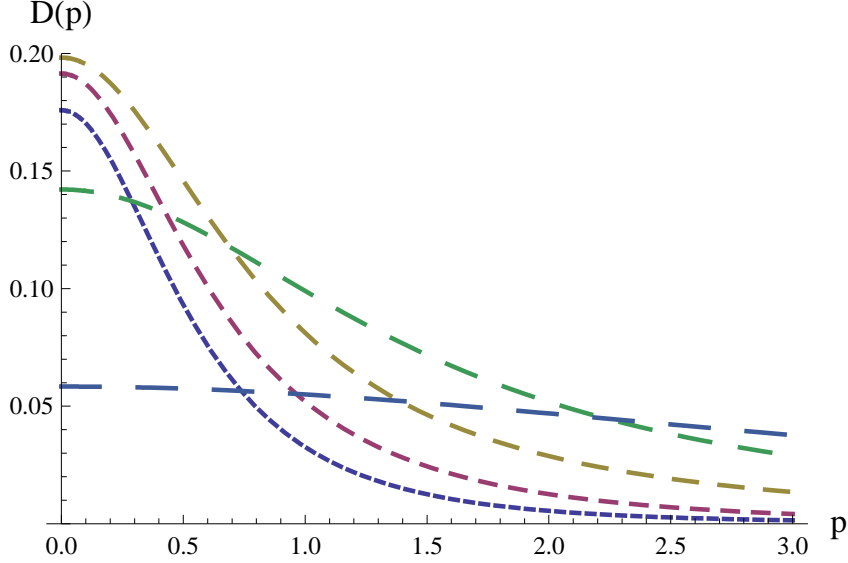}
\end{center}
\caption{The quark mass function difference ${\cal D}(p)=m(p)-m_0$ 
measuring the extent of dynamical mass generation, is
represented with increasing number of dashes per curve 
for five different current quark masses $m_0$
with values $10^{-4}, 10^{-1}, 10^{-1/2}, 10^{0}, 10^{1/2}$, 
in dimensionless units of $\sigma=0.19$ GeV$^2=1$.
\label{fig_massinsomecases}}
\end{figure}

Unlike the fixed point method, converging quite fast (with a single iteration) for large current quark masses $m_0 >> \sqrt \sigma $, 
the variational method converges faster for small current quark masses. Thus both methods  
are complementary. In Tables \ref{table 1} and \ref{table 2}
and in Fig. \ref{fig_solutionmodel4} 
we show the results of our minimization for two ansatze and for different current quark masses spanning over
five orders of magnitude. In Figs. \ref{fig_massgap} and \ref{fig_massinsomecases} we illustrate the
mass difference at the momentum origin $m(0)-m_0$, the mass gap $m(0)$ and the running mass $m(p)$ 
for different current masses $m_0$.

\C{
Now that we have an excellent and simple ansatz for the running quark mass, we may compute the
regularized quark condensate and the quark dispersion relation. The quark condensate $\langle \bar \psi \psi \rangle$
is another possible order parameter, to be compared with the other order parameter
we computed, {\it i. e. } the quark mass at vanishing momentum $m(0)$.
The quark condensate is computed from the one-loop quark propagator functions in Eq. (\ref{AB}), and
it is ultraviolet divergent for finite quark masses. Thus we regularize the quark condensate, subtracting 
the quark condensate for a current quark,
\be
\langle \bar \psi \psi \rangle -\langle \bar \psi \psi \rangle_0
=  -{g \over  \pi^2}
\int_0^\infty k^2 \, d k   \left[ {m (k)  \over \sqrt{k^2 + m(k)^2}} -  {m_0  \over \sqrt{k^2 + {m_0}^2}} \right]
\ee
}

\C{
The one quark dispersion relation $E(p)$, defined in Eq. (\ref{quark energy}), 
is relevant for the boundstate equation of mesons or of baryons.
For instance, in the instantaneous Salpeter equation, 
a hamiltonian $H= E_q+ E_{\bar q} + V_{q \, \bar q}$ can be defined for mesons
(actually the hamiltonian is a matrix 
\cite{Bicudo_scapuz}
including negative and positive energy components).
The dispersion relation $E(p)$
is infrared divergent due to the infinite constant of quark-antiquark potential detailed 
in Eq. (\ref{IRdivpot}).  In momentum space, this leads to an infinite Dirac delta
in the integral present in $E(p)$. We can regularize the numerical integral, subtracting
a term to cancel the integrand when $k=p$, a term that we add back analytically, 
\bea
&& E(p) =  
p C(p) + m_0 S(p)  + {\sigma \over p^2}\, \int_0^\infty {d k \over 2 \pi}   \biggl\{
I_A(k,p,\mu) \,   \left[ S(k) - S(p) \right ] S(p) 
\label{regularized E}
\\ 
\nonumber 
&& +  I_B(k,p,\mu) \,   \left[ C(p) -C(p) \right] C(k)  + \left[ I_A(k,p,\mu) S^2(p)+ I_B(k,p,\mu) C^2(p) \right] \biggr\} \ ,
\eea
in particular the integral of $I_A \,  S^2+I_B \, C^2$ is  analytical thanks to Eq. (\ref{analytical A B}),
and in the limit of a vanishing infrared regulator $\mu \to 0$ this term, that we subtracted and must now
add back to the quark energy, reduces to a simple infrared divergence plus a finite term,
\be
{\sigma \over p^2}\, \int_0^\infty {d k \over 2 \pi}  I_A(k,p,\mu) S^2(p)+ I_B(k,p,\mu) C^2(p) 
\to
{\sigma \over 2 \mu} - { 2 \sigma \over \pi} { p \over  p^2 - m(p)^2 } 
\ .
\label{analytical regulator}
\ee
\C{
\begin{table}[t!]
\begin{center}
\begin{tabular}{c| c c c c c}
\hline 
$m_0$ & $  \left(-\langle \bar \psi \psi \rangle +\langle \bar \psi \psi \rangle_0 \over g\right)^{1\over 4} $ & $e_0$ & $e_1$ & $e_3$ & $e_5$ \\
\hline
 $ 0.000100000 $ & $ 0.255569 $ & $ 0.21104 $ & $ 1.9549 $ & $ 102.65 $ & $ 451.90 $  \\ 
 $ 0.000316228 $ & $ 0.255847 $ & $ 0.21156 $ & $ 1.9539 $ & $ 102.23 $ & $ 448.34 $  \\ 
 $ 0.00100000  $ & $ 0.256701 $ & $ 0.21321 $ & $ 1.9518 $ & $ 100.94 $  & $ 436.59 $ \\ 
 $ 0.00316228  $ & $ 0.259029 $ & $ 0.21814 $ & $ 1.9488 $ & $ 97.304 $  & $ 399.80 $ \\ 
 $ 0.0100000    $ & $ 0.267020 $ & $ 0.23387 $ & $ 1.9269 $ & $ 86.717 $  & $ 317.83 $ \\ 
 $ 0.0316228    $ & $ 0.287014 $ & $ 0.27830 $ & $ 1.8859 $ & $ 65.176 $  & $ 173.48 $ \\ 
 $ 0.100000      $ & $ 0.331664 $ & $ 0.39429 $ & $ 1.8758 $ & $ 35.484 $  & $ 49.785 $\\ 
 $ 0.316228      $ & $ 0.428943 $ & $ 0.71809 $ & $ 1.8628 $  & $ 13.582 $ & $ 3.6035 $ \\
 $ 1.00000       $ & $ 0.771803 $ & $ 1.6210 $ & $ 2.2704 $  & $ 2.8947 $ & $ 0.0076 $  \\ 
 $ 3.16228       $ & $ 0.984801 $ & $ 4.7134 $ & $ 2.4773 $ & $ 0.34516 $  & $ 0.0000 $ \\ 
 $ 10.0000       $ & $ 1.26682 $ & $ 14.999 $ & $ 2.4199 $ & $  0.03598$  & $ 0.0000 $ \\ 
 $ 31.6228       $ & $ 1.60590 $ & $ 48.113 $ & $ 2.3283 $ & $ 0.00371 $  & $ 0.0000 $ \\ 
\hline
\end{tabular}
\caption{The results for the regularized quark condensate and for the
regularized quark dispersion relation obtained with the ansatz of Eq. (\ref{quarkenergy}). 
All results are in dimensionless units of $\sigma=0.19$ GeV$^2=1$.
\label{table 3} }
\end{center}
\vspace{0.5cm}
\end{table}
}
Importantly, this divergence is physically irrelevant since, in the hamiltonian of any colour singlet 
hadron, the sum of the divergences of the quark and of the antiquark energies  cancel
with the infrared divergence of the quark-antiquark potential detailed in Eq. (\ref{IRdivpot}). 
Finally for the purpose of future computations of the hadron spectra, it is convenient
to write the dispersion relation as a sum of the analytical infrared term  of Eq. (\ref{analytical regulator}), 
plus the free quark dispersion relation dominating the ultraviolet,
and plus a finite and compact term $\widetilde E(p)$, an integral that we compute numerically,
\bea
E(p) & = & {\sigma \over 2 \mu} - { 2 \sigma \over \pi} { p \over  p^2 - m(p)^2 } + p C(p) + m_0 S(p)  + \tilde E(p)  \ .
\label{compact energy}
\eea
The numerical integral $\widetilde E(p)$ decays like $1 /p^5$ in the chiral limit of small current quark
masses and decays like $1/p^3$ in the case of large current quark masses. $\widetilde E(p)$ is negative
and we can conveniently fit it with the rational function, or Pad\' e approximant with odd powers of $p$ only,
\be
\widetilde E(p) \simeq -{1 \over e_0 +e_1 p + e_3 p^3 + e_5 p^5} \ .
\label{quarkenergy}
\ee
We show the best fitting parameters $e_0, e_1, e_3, e_5$ in Table \ref{table 3}. With our
excellent fits of  the dynamical quark mass $m(p)$ and of the quark dispersion relation $E(p)$ 
we are well equipped to address further problems, 
such as the breaking of chiral symmetry or the hadronic excited spectra at finite temperature $T$.
}

\section{Conclusion}

While the chiral limit of
$m_0 <<  \sqrt \sigma $ was already well known in the literature, we find unanticipated  effects for finite $m_0 \simeq
\sqrt \sigma$ and for heavy $m_0  >>  \sqrt \sigma$ current quark masses.
We study in detail the large $m_0$ limit, performing an one-loop expansion in the dimensionless 
number $\sigma / {m_0 }^2$.
We also develop a new technical approach to solve the mass gap equation, utilizing the variational
principle to increase the precision of our mass solution $m(p)$ in the infrared limit of 
$p \to 0$, relevant to compute the mass gap $m(0)$, an order parameter for the chiral
phase transition. 
We also show that the dynamical generated constituent quark mass $m(p)$ can be quite well fitted by 
our inverse even quartic polynomial ansatz, 
\C{
a Pad\' e approximant   
}
with parameters $c_0, \, c_2$ and $c_4$ depending only on the
current mass $m_0$.

\begin{figure}[t!]
\includegraphics[width=.55\columnwidth]{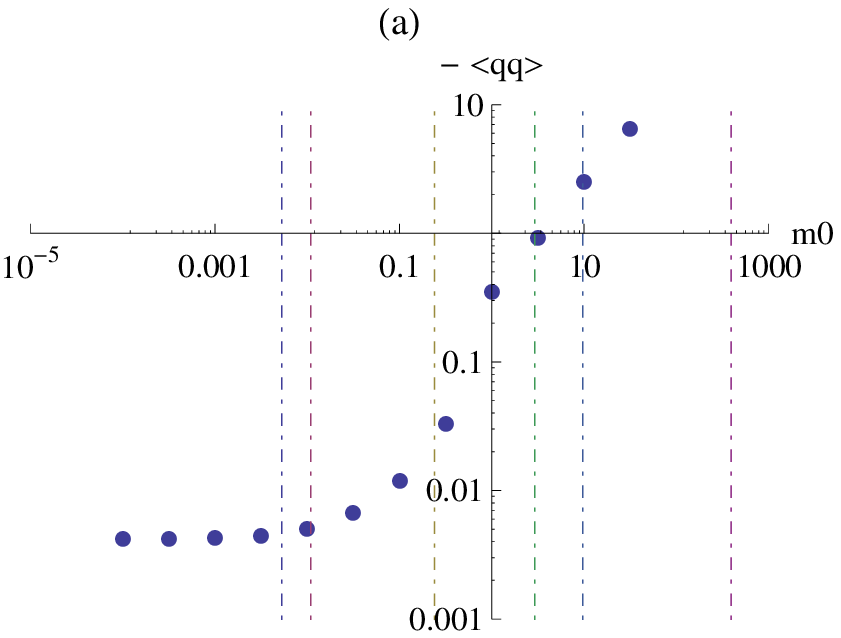}
\hspace{0.1cm}
\includegraphics[width=.45\columnwidth]{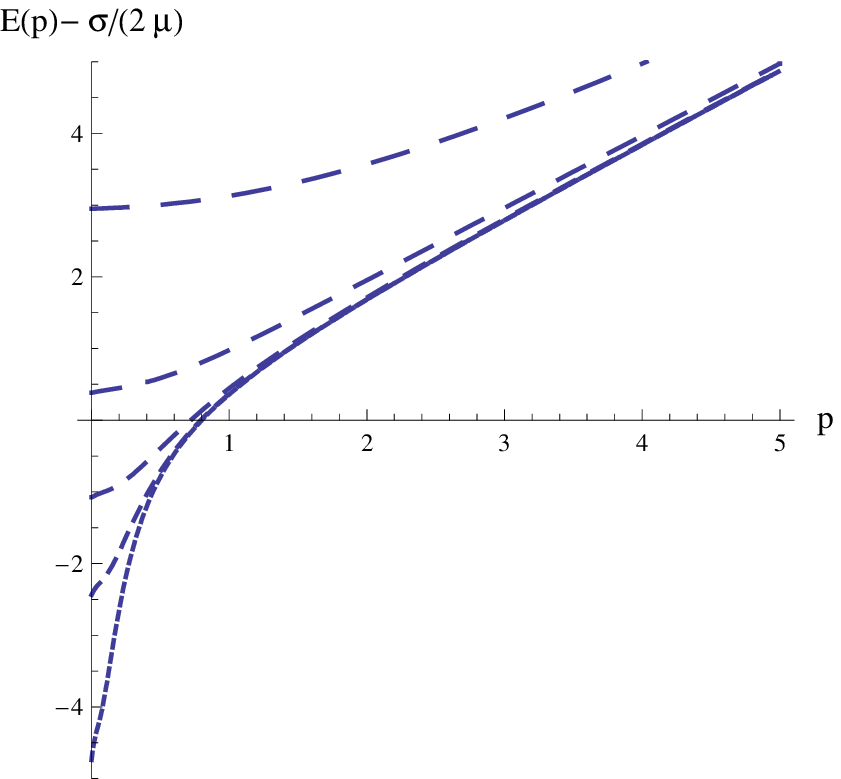}
\caption{ 
(a) We show a log log plot of minus the regularized quark condensate $- \langle \bar \psi \psi \rangle +  \langle \bar \psi \psi \rangle_0$
as a function of the current quark mass $m_0$, with vertical dot-dashed lines representing the
current masses of the quarks $u$, $d$, $s$, $c$, $b$, $t$.
(b) We represent the regularized quark dispersion relation $E(p) - {\sigma \over 2  \mu}$ 
with increasing number of dashes per curve 
for five different current quark masses $m_0$
with respective values $10^{-4}, 10^{-1}, 10^{-1/2}, 10^{0}, 10^{1/2}$.
The quark condensate has an inflection point 
for finite quark masses close to the strange quark mass, and 
for large masses it rises linearly with $m_0$.
All results are in dimensionless units of $\sigma=0.19$ GeV$^2=1$. 
\label{solutionmodel4}}
\end{figure}

Our surprising results are that the dynamical mass generation has finite effects persistent beyond
the chiral limit. At $m_0 \simeq \sqrt \sigma$ , in particular for masses similar to the strange quark
$s$ mass, the quark mass generation  $ m(0) -m_0$ is maximum,
as shown in Fig. \ref{fig_massinsomecases}, 
while one would naively expect the quark mass
generation to be maximum in the chiral limit close to the up $u$ or down $d$ quark masses. 
A second order parameter, 
the regularized quark condensate $\langle \bar \psi \psi \rangle -  \langle \bar \psi \psi \rangle_0$,
monotonously grows in absolute value with $m_0$ 
and shows an inflexion point for masses similar to the strange quark $s$ mass,
as depicted in Fig. \ref{solutionmodel4}. 
In the limit of heavy quark masses  $m_0  >>  \sqrt \sigma$ of the charm $c$, bottom $b$ and top $t$ quarks,   
although the mass gap difference $ m(0) -m_0$ vanishes like $ \sigma / m_0$, it occurs that the
energy density difference  ${\cal E} - {\cal E}_0$ is maximum. In the heavy quark limit, the energy
density difference converges to a constant limit when $m_0 \to \infty$, 
and we show in Fig. \ref{fig_solutionmodel4} that it is one order of magnitude larger than in the chiral limit. 
This may be relevant for cosmology, contributing to the dark energy.

\C{
The numerical variational technique developed here, together with the detailed 
solutions for the running quark mass $m(p)$ and for the quark dispersion relation $E(p)$
as a function of the current quark mass $m_0$ and of the string tension
$\sigma$, are necessary tools for the continuation of our program to study the QCD 
phase diagram and the hadron spectrum at finite temperature $T$ and density $ \mu$, 
when the string tension $\sigma$  becomes quite small, possibly smaller than
the light current quark masses.
}

{\bf acknowledgements}
\\
I am very grateful to Gast\~ao Krein on the variational method, 
and to Marlene Nahrgang, to Pedro Sacramento and to Jan Pawlowski 
for discussions on the QCD phase diagram motivating this paper. 
I acknowledge the financial support
of the FCT grants CFTP, 
CERN/ FP/ 109327/ 2009 
and
CERN/ FP/ 109307/ 2009.


\end{document}